\documentclass[pra,twocolumn,amsfonts,amssymb,amsmath,floatfix,tightenlines,superscriptaddress]{revtex4-1}
\usepackage{amsfonts,amssymb}
\usepackage[subnum]{cases}
\usepackage{mathrsfs}
\usepackage[tone]{tipa}
\usepackage{amsmath}
\usepackage[none]{hyphenat}
\usepackage{graphicx}
\usepackage{hyperref}    
\usepackage{footmisc}
\usepackage{bm}          
\usepackage{natbib}
\usepackage{color}
\usepackage{longtable}
\usepackage{ textcomp }
\usepackage{verbatim}

\begin{document}

\title{
Dynamical preparation of an atomic condensate in a Hofstadter band}

\author{Han Fu}
\affiliation{James Franck Institute, University of Chicago, Chicago, IL 60637, USA}
\affiliation{Department of Physics, University of Chicago, Chicago, IL 60637, USA}
\author{Andreas Glatz}
\affiliation{Materials Science Division, Argonne National Laboratory, Argonne, IL 60439, USA}
\affiliation{Department of Physics, Northern Illinois University, DeKalb, IL 60115, USA}
\author{F. Setiawan}
\affiliation{Pritzker School of Molecular Engineering, University of Chicago, Chicago, IL 60637, USA}
\affiliation{James Franck Institute, University of Chicago, Chicago, IL 60637, USA}
\affiliation{Department of Physics, University of Chicago, Chicago, IL 60637, USA}
\author{Kai-Xuan Yao}
\affiliation{James Franck Institute, University of Chicago, Chicago, IL 60637, USA}
\affiliation{Department of Physics, University of Chicago, Chicago, IL 60637, USA}
\affiliation{Enrico Fermi Institute, University of Chicago, Chicago, IL 60637, USA}
\author{Zhendong Zhang}
\affiliation{James Franck Institute, University of Chicago, Chicago, IL 60637, USA}
\affiliation{Department of Physics, University of Chicago, Chicago, IL 60637, USA}
\affiliation{Enrico Fermi Institute, University of Chicago, Chicago, IL 60637, USA}
\author{Cheng Chin}
\affiliation{James Franck Institute, University of Chicago, Chicago, IL 60637, USA}
\affiliation{Department of Physics, University of Chicago, Chicago, IL 60637, USA}
\affiliation{Enrico Fermi Institute, University of Chicago, Chicago, IL 60637, USA}
\author{K. Levin}
\affiliation{James Franck Institute, University of Chicago, Chicago, IL 60637, USA}
\affiliation{Department of Physics, University of Chicago, Chicago, IL 60637, USA}
\date{\today}

\begin{abstract}
The creation of a Hamiltonian in the quantum regime which has non-trivial
topological features is a central goal of the
cold-atom community, enabling widespread exploration of novel phases
of quantum matter.
A general scheme to synthesize such Hamiltonians is based on dynamical
modulation of optical lattices which thereby generate vector potentials.
At the same time the modulation can lead to heating and
serious difficulties with equilibration.
Here we show that these challenges can be overcome by demonstrating how a
Hofstadter Bose-Einstein condensate (BEC) can be dynamically realized,
using experimental protocols. From Gross-Pitaevskii simulations
our study reveals a complex, multistage evolution; this includes a chaotic intermediate ``heating" stage followed by a spontaneous reentrance to the BEC.
The observed behavior is reminiscent of evolution in cosmological models.
\end{abstract}

\maketitle

\section{Introduction}

One of the great challenges in the field of ultracold atoms is to realize
a topological phase of a quantum, many-body system.
While a number
of novel Hamiltonians have
been realized experimentally \cite{Esslinger,Bloch,Ketterle}, often based on artificial
gauge fields
\cite{Spielman,Eckardt,Goldman_2014,physicsToday,Dalibard_2011,Struck,Dalibard_2014,Dalibard_2015,Monika_2018,Demler},
observing collective physics has remained challenging.
Creating a Hamiltonian with nontrivial topological properties such as the iconic
Harper-Hofstadter model
\cite{TKNN}
and addressing it within the quantum regime will enable wide ranging explorations of topological phases \cite{Kane,SCZ,Kane2}. This has
implications for atomic and condensed matter physics and other sub-disciplines as well.

Almost all schemes for arriving at these Hamiltonians
are dynamical in nature
\cite{Dalibard_2014,Sengstock,Lewenstein,Polkovnikov,Polkovnikov2}.
In cold atomic gases they involve
the introduction of time-dependent optical lattices which generate artificial gauge fields.
Unfortunately, this dynamical engineering
has an important adverse consequence: heating \cite{Cooper,Bloch2}, which presents
impediments for reaching the quantum regime.
Successfully implementing
this time-dependent or ``Floquet" engineering in the quantum regime is, hence, a central goal of our larger
community.
Most urgent is to identify
the pathways involved in these successful realizations.
Importantly, there are no known fundamental barriers for arriving in the quantum regime of
the classic Hofstadter model. The MIT group\cite{Ketterle2} has reported evidence
for a Bose condensate in a Hofstadter band.
The Munich group \cite{Bloch}, which
has  simulated this Hamiltonian and observed
topological features \cite{Bloch}, has, however, met difficulties in reaching the ground state.

It should be emphasized that Floquet
generation of topological bands involves multi-band participation with
band crossing and inversion, abrupt band minima transitions and complex patterns
of flux penetration.
In more conventional, non-topological systems
\cite{Eckardt} in which new band minima are
created by Floquet engineering the changes in the bandstructure are more continuous. This
derives from the fact that
a finite critical shaking
amplitude
is required to obtain new band minima, which, in turn
controls the onset of the equilibration process.
In the Hofstadter case, by contrast,
an infinitesimally small shaking
amplitude will shift the ground state minima and abruptly initiate
equilibration. As a consequence we find that the resulting dynamics
leads to a chaotic, intermediate heating stage concomitant with the introduction of
flux and the related reorganization of the complex condensate phase pattern. The
novel dynamics
makes
BEC formation in Floquet-engineered topological systems both richer and more complicated.

A central goal of this paper is to
elucidate how such a superfluid can be successfully realized given these complications.
This involves
characterizing
the requisite dynamical pathways.
Instrumental to our paper is, then, the time evolution.
Here we emphasize our rather unexpected finding
involving highly chaotic behavior at intermediate times en route to
re-condensation in a topological band.
We emphasize that understanding
these phenomena should prove an enormous benefit
to the cold-atom, as well as to the solid-state
\cite{Fsolid,Hunt,Kim,Geim}
and photonics
communities \cite{Jon}
with a shared interest in
Floquet engineering.

In our theoretical investigation into the time sequence involved
in Hofstadter band condensation, we focus on three challenges which the system must address.
First, there is heating from the direct application of the Floquet drive.
A second challenge arises from the sudden change of the many-body ground state. Meeting this challenge requires that
the condensate wavefunction quickly develop a specific and complex phase pattern.
A third challenge comes from accommodating interparticle interaction effects which are required
for equilibration, but not generally compatible with analytical predictions based on Floquet
engineering.

\begin{figure}[h]
\includegraphics[width=.5\textwidth]
{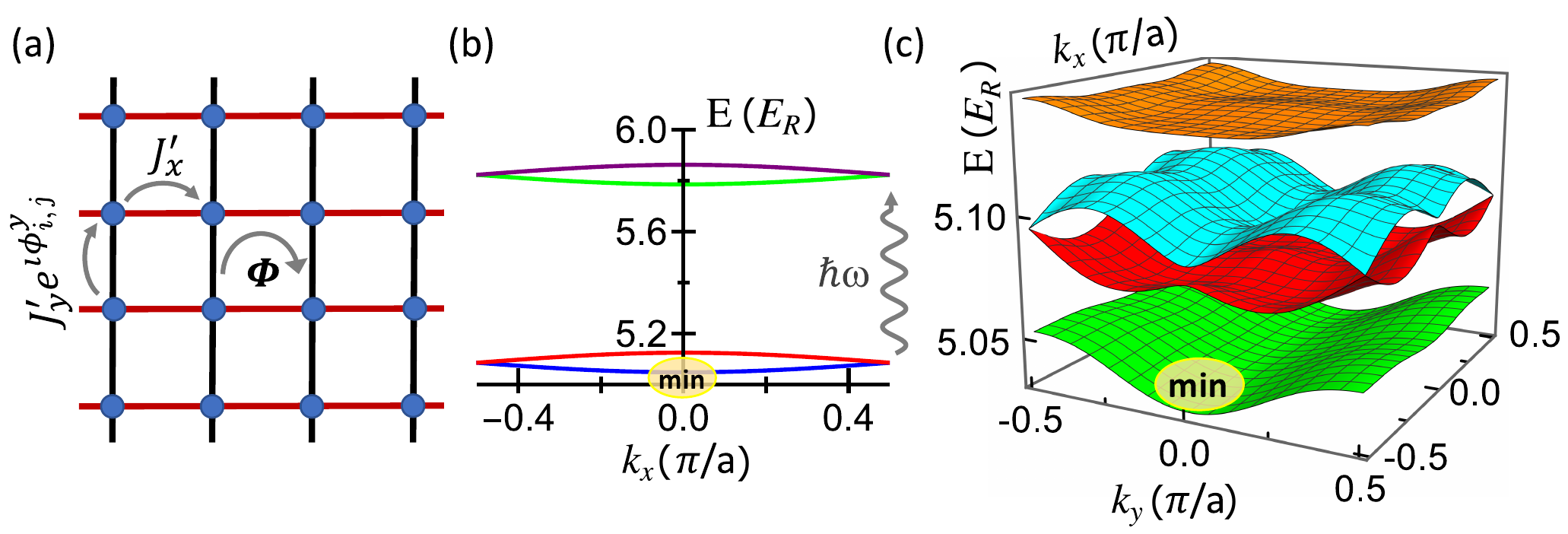}
\caption{Characteristics of the Floquet-engineered Hofstadter model. (a)~Illustration of the
ideal Hofstadter Hamiltonian which is the target model, only approximately realized through
a lattice-shaking protocol
based on Eq.~\eqref{eq:V} which is used here.
$J_x^\prime,\,J_y^\prime$ are the $x$ and $y$ axis tunneling parameters,
and the phase
$\phi^y_{i,j}=2\alpha \pi i+\frac{\pi}{2}j$.
The effective flux $\Phi$ inside each square cell is $1/4$ times the flux quantum.
(b)~Pre-shaking energy bandstructure at $k_y=0$ based on
Eq.~\eqref{eq:V}, where
the energy $E$ is in recoil units $E_R$. Here $\omega$
is the modulation frequency which couples the bands. Unless noted otherwise, throughout the paper we use
\cite{Bloch,Cooper},
$V_y=6 E_R,\,V_x=10E_R,\,V_{yl}=0.81E_R$, with shaking frequency $\omega=0.72E_R/\hbar$, where the recoil energy $E_R=\hbar^2(\pi/a)^2/2m$, and $a$ is the lattice constant of the underlying square lattice, and $m$ is the atomic mass.
(c)~Floquet-engineered Hofstadter bands
at $\kappa=0.58\hbar\omega$. The energy separation between states ${\bf{k}}=0$ and ${\bf{k}}=(\pm\pi/2a,0)$ is roughly $0.0025 E_R$.
Here and throughout the paper, the ground state of this Floquet-engineered Hofstadter bandstructure
is indicated by `min'.
}
\label{fig:bandcross}
\end{figure}

Our paper reports a rich set of dynamical
processes en route to forming a BEC in a Floquet-Hofstadter bandstructure.
These observations are derived from Gross-Pitaevskii (GP) simulations.
Despite the aforementioned challenges, we are able to provide a large body of evidence supporting the emergence
of condensation in a Hofstadter lattice.
Moreover, our analysis shows that a substantial fraction of the atoms are in
the ground state.
Interestingly we observe a multistage dynamics, which has features in common with
models of cosmological evolution \cite{reheating}. This starts with a coherent series of oscillations of the population
and
is followed by
a chaotic ``heating" stage, which is accompanied by an
abrupt injection of magnetic flux \cite{Appendix,Spielmanvortices}.
We assume no dissipation in our simulations but, surprisingly observe
that the system is ultimately able to spontaneously relax into the new ground state where condensation occurs.
The intermediate heating stage, in particular,
is found to be essential, enabling reentrance to this new condensate.
When the system reaches the steady state, we are able to
extract occupations of different Hofstadter bands, thus characterizing a small number of
excitations which coexist with the BEC in the ground state.

\section{Floquet-Hofstadter theory}

We follow the approach used by the Munich group \cite{Bloch,Cooper} for Floquet engineering of the Hofstadter Hamiltonian.
This involves loading bosons into a two-dimensional optical lattice with a potential which includes both
a periodically oscillating contribution $V_{os}$, and a static superlattice  $V_{st}$. We consider
a square lattice having lattice constant $a$ superposed on an additional lattice with constant $2a$ in the $y$
direction. The lattice potentials are given by
\begin{equation}
\begin{aligned}\label{eq:V}
V_{st}=&V_x \sin^2\left(
\pi x/a\right)
+ V_y \sin^2\left(\pi y/a\right)
+ V_{yl}\sin^2\left(
\pi y/2a\right),
\\
V_{os}=&\kappa \left[\sin\left(\pi/4+\pi y/2a\right) \cos\left(\phi_0 + \omega t-2\alpha \pi x/a\right)\right.\\
&\left.+\cos\left(\pi/4+\pi y/2a\right) \sin\left(\phi_0-\omega t-2\alpha \pi x/a\right)\right],
\end{aligned}\end{equation}
where $V_x, \,V_y,\,V_{yl}$ represent the strengths of the respective components in the static lattice, $\kappa$ is the amplitude of the oscillating lattice which is identically zero before we turn on the shaking and $\phi_0 = \pi/4$. The
presence of a time-dependent potential $V_{os}$ enables the atoms to tunnel in the $y$ direction
and acquire a position-dependent, Aharonov-Bohm-like phase. Here $\alpha$ is the ratio between the flux per unit cell of the square lattice $\Phi$ to the flux quantum $\Phi_0$, with $\Phi/\Phi_0\equiv\alpha=1/4$. Our GP simulations are based on a Hamiltonian which
includes two body interactions and the single particle contributions from both kinetic energy
terms and lattice potentials which directly implement Eq. \eqref{eq:V}. We consider a two-dimensional system.

\begin{figure}[h]
\includegraphics[width=.47\textwidth]
{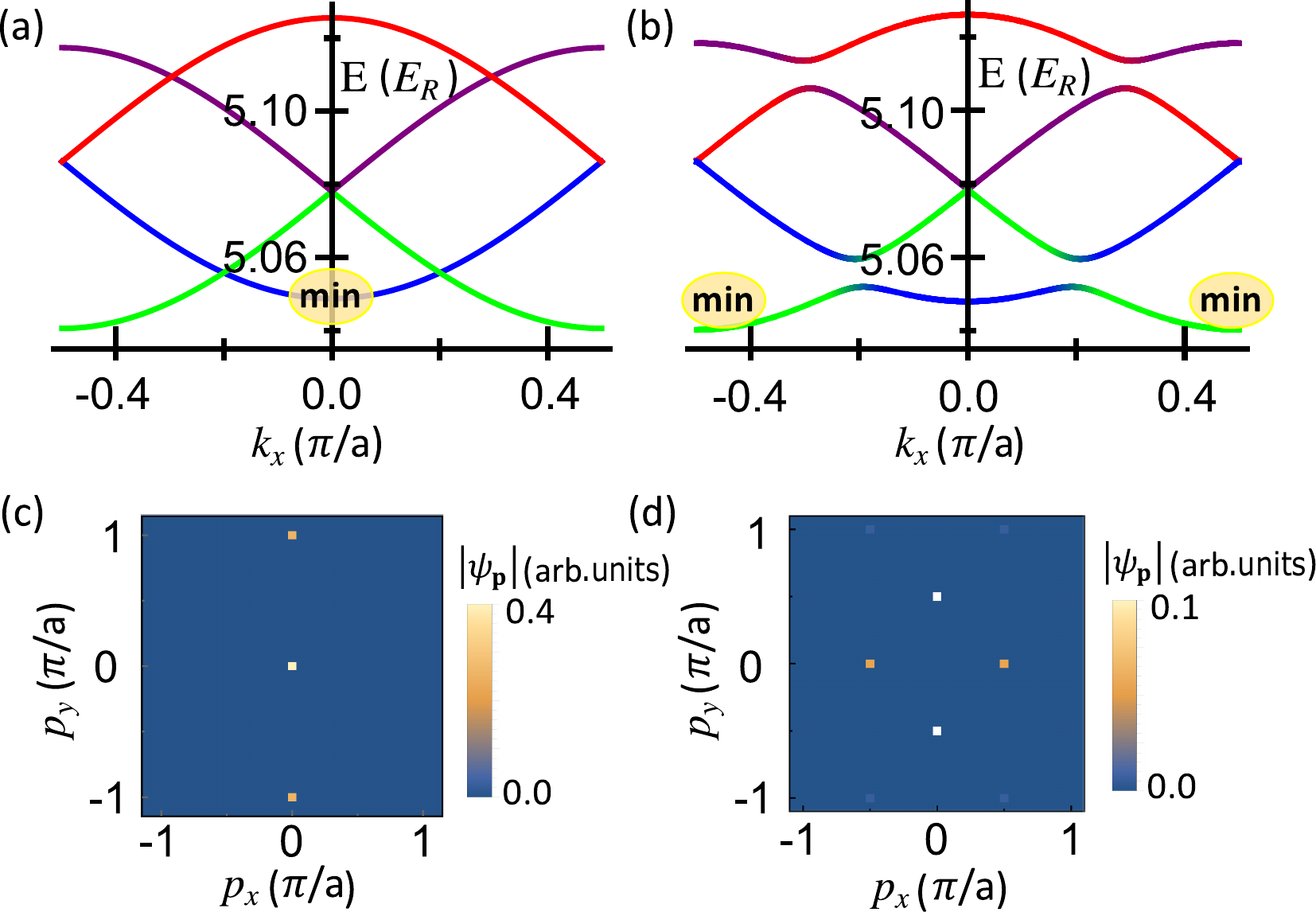}
\caption{ Abrupt changes in energy bands and wavefunctions due to shaking, obtained from
numerical calculations using Floquet theory. (a)~Energy bands at $k_y=0$ without shaking; $\bf{k}$ is the magnetic zone quasi-momentum.
This figure is identical to Fig.~\ref{fig:bandcross}(b), now replotted in the framework of Floquet theory. (b)~Energy bands at
$\kappa=0.1\hbar\omega$. Panels (c) and (d) respectively show
ground-state wavefunctions
in the momentum ($\bf{p}$) space
at $\kappa=0$ and $0.1\hbar\omega$. These are labeled as `min' in the
Floquet-Hofstadter bands in (a) and (b).
The abrupt change of characteristic momenta as $\kappa$ varies reflects a first-order-like transition.
Color codes indicate contribution from the initial bands to the Floquet-Hofstadter bands; intermediate colors in (b) represent band hybridization.
Note that Fig.~\ref{fig:bandcross}(c) and Fig.~\ref{fig:wave}(b)
represent slightly different parameter sets, with the latter chosen for pedagogical purposes to illustrate
more clearly the dramatic change in the band structure that ensues even at very small $\kappa$.
}
\label{fig:wave}
\end{figure}

When the modulation energy $\hbar\omega$ is much larger than the effective tunneling parameters, the system approaches the ideal Hofstadter
Hamiltonian
\cite{Polkovnikov2,Monika_thesis,thomas_thesis}. This Hamiltonian has only nearest-neighbor tunneling
in the $x$- and $y$-direction denoted by $J^\prime_{x}$
and
$J^\prime_{y}
e^{\imath\phi^y_{i,j}}$ respectively,
where
$\phi^y_{i,j}=2\alpha \pi i+\frac{\pi}{2}j$ and $J_x^\prime,\,J_y^\prime$ are real. The coordinates here are $(x,y) = (i,j)a$.
This ideal case, schematically illustrated
in Fig.~\ref{fig:bandcross}(a), should be contrasted with the Floquet-Hofstadter realization based on
Eq.~\eqref{eq:V}. Our theory implements the full dynamical Hamiltonian, which naturally includes higher-order terms in $J^\prime_x/\hbar\omega$ and $J^\prime_y/\hbar\omega$.

We characterize this latter Floquet-Hofstadter Hamiltonian through the
resulting band structure. In the absence of shaking
($\kappa =0$), the band structure
obtained from Eq.~\eqref{eq:V}, is
shown in Fig.~\ref{fig:bandcross}(b).
By contrast
when $\kappa$ assumes the experimental value
\cite{Bloch,Cooper} ($\kappa=0.58\hbar\omega)$ a very different band dispersion emerges which is
presented
in Fig.~\ref{fig:bandcross}(c).

It should be stressed that once a shaking amplitude $\kappa \neq 0$
is applied, regardless of how small $\kappa$ is, there is a dramatic change of the ground state. 
We contrast the bandstructures for the two situations: in the absence of shaking [Fig.~\ref{fig:wave}(a)]
and at a small shaking amplitude $\kappa=0.1\hbar\omega$ [Fig.~\ref{fig:wave}(b)]. Here
the energy minima in the ground band
shift their
position abruptly from the band center
${\bf{k}}=0$
to the band edge ${\bf{k}}=(\pm \frac{\pi}{2a},0)$. Note that, these two quasi-momenta ${\bf{k}}=(\pm\frac{\pi}{2a},0)$ are connected by a reciprocal vector and thus correspond to a unique state; this is henceforth called the
``Floquet-Hofstadter ground state" \footnote{The Floquet-Hofstadter ground state is defined as the lowest energy state in the ground band. This is the band in which the ${\bf{k}}=0$ state is adiabatically connected to the ground state in the static system.} .

The ground-state wavefunctions also exhibit a
discontinuous change as can be seen by
comparing their behavior without shaking and with shaking
at $\kappa=0.1\hbar\omega$
through their distribution in momentum space, see Figs.~\ref{fig:wave}(c) and
\ref{fig:wave}(d), respectively.
In these figures
the wavefunctions are expanded in terms of
${\bf{p}}={\bf{k}} +{\bf{G}}$, where $\bf{G}$ are
the reciprocal wavevectors of the oscillating lattice potential.
Before shaking, the atoms are confined to ${\bf{p}} =(0,0)$. (The two extra spots appearing
in
Fig.~\ref{fig:wave}(c) are associated with higher reciprocal vectors.)
In the presence of lattice shaking a new set
of four characteristic momenta emerge, represented by ${{\bf{p}}  = (\pm\frac{\pi}{2a}, 0),\,(0,\pm \frac{\pi}{2a}})$.
Importantly, a macroscopic population of these four momentum states can serve as a signature that particles are occupying the ground state of the Floquet-Hofstadter band.

We simulate the dynamics of the atoms in the Floquet lattice through a GP numerical procedure which
uses a Graphics Processing Unit-based
quasispectral, split-step method to solve the GP equation based on fast Fourier
transforms \cite{Andreas}.
Here we
include
a small, nonzero interparticle interaction potential $U_0= 7.5 \times 10^{-4}  E_R$.
In our numerical simulations, we start with a condensate in
the static lattice $V_{st}$ and linearly ramp up the shaking amplitude $\kappa$ in $V_{os}$ of Eq.~\eqref{eq:V}.
After the ramp,
$\kappa$ is held constant; from the GP simulations we
are able to examine the full evolution of the time-dependent wavefunction in both real and momentum space.

\begin{figure}[h]
\includegraphics[width=0.45\textwidth]
{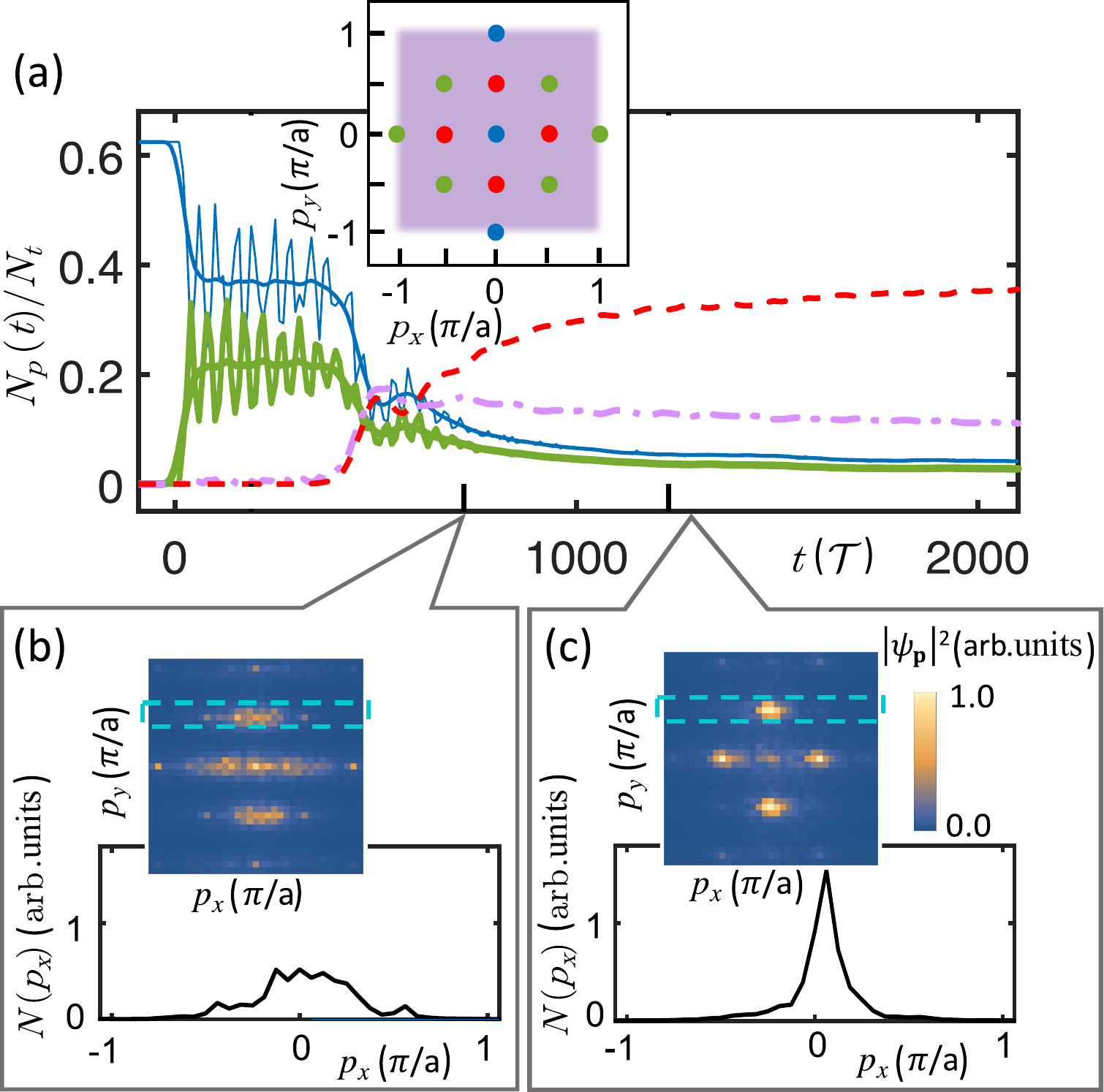}
\caption{Quench dynamics of Bose condensates undergoing transition from conventional bands (with $\kappa=0$) to Hofstadter bands (at $\kappa=0.58\hbar\omega$)
with $\kappa$ linearly ramped from zero to $0.58\hbar\omega$ within $30\,\mathcal{T}$ where $\mathcal{T}$ is the Floquet period.
(a)~Time dependence of particle populations in four characteristic momentum groups labeled using the color code in the
inset: dashed line (red), dash-dotted line (purple), thick solid line (green), and thin solid line (blue). The transfer of boson populations between different groups indicates a three-stage evolution. After initial oscillations in the first stage, a `heating' state emerges which then spontaneously transitions to the final condensation stage.
(b)~Particle population at $p_y=\pi/2a$ for $t=720\,\mathcal{T}$ in the `heating' stage. The inset is the corresponding image in the full momentum space within the same $2\pi/a\times2\pi/a$ Brillouin zone as in the inset of (a), where the blue dashed box indicates the relevant vertically integrated region. (c)~Counterpart of (b) at $t=1230\,\mathcal{T}$ in the condensation stage.
The transition from broad distribution in (b) to sharp peaks along the $p_x$ direction in (c) provides some evidence for condensate formation.
This analysis shows that a sizeable (about $40 \%$) fraction of the atoms is condensed.
}
\label{fig:ktime}
\end{figure}

\section{Characterizing the evolutionary pathways}

Our simulations reveal a rich dynamics when the atoms transfer to the Floquet-Hofstadter band.  We observe three distinct evolutionary stages, as we follow the momentum-space populations, see Fig.~\ref{fig:ktime}(a). Below we outline the key features of each stage.
In the first stage (from $0$ to $300\,\mathcal{T}$), we see a period of coherent oscillations which involves
transient occupations of higher bands.
A complicated dynamics then ensues within the second
stage (from $300$ to $900\,\mathcal{T}$). Here the
population becomes widely distributed over different states and different bands, see Fig.~\ref{fig:ktime}(b). We
refer to this second stage as the ``intermediate
heating"
stage, where interesting, highly chaotic behavior occurs. This
time period
reflects the non-adiabatic evolution and is reminiscent of the
``preheating" and ``turbulent" stages associated with inflationary models of
cosmology \cite{reheating,reheating2,reheating3}.
It is during this second stage, as the wavefunction begins to develop a new and complex phase pattern,
that we observe a sudden onset of flux penetration \cite{Appendix}.

By contrast, in the third stage (beginning around $900\,\mathcal{T}$), the population starts to settle into the Floquet-Hofstadter ground state.
This appearance of population accumulation into the ground state is suggestive of Bose condensation.
One sees
that a rather sharp momentum
distribution emerges during this time, see, for example, Fig.~\ref{fig:ktime}(c).
This matches that of the ground state shown in Fig.~\ref{fig:wave}(d).
We emphasize that the evolution occurs spontaneously in our simulations \cite{Appendix}, which are calculated without dissipation (consistent
with experimental conditions).
Notably, this transition into the final ground state is only possible in the presence of two-body interactions which drive collisions and subsequent relaxation into a new set of momenta.

\begin{figure}[h]
\includegraphics[width=.45\textwidth]
{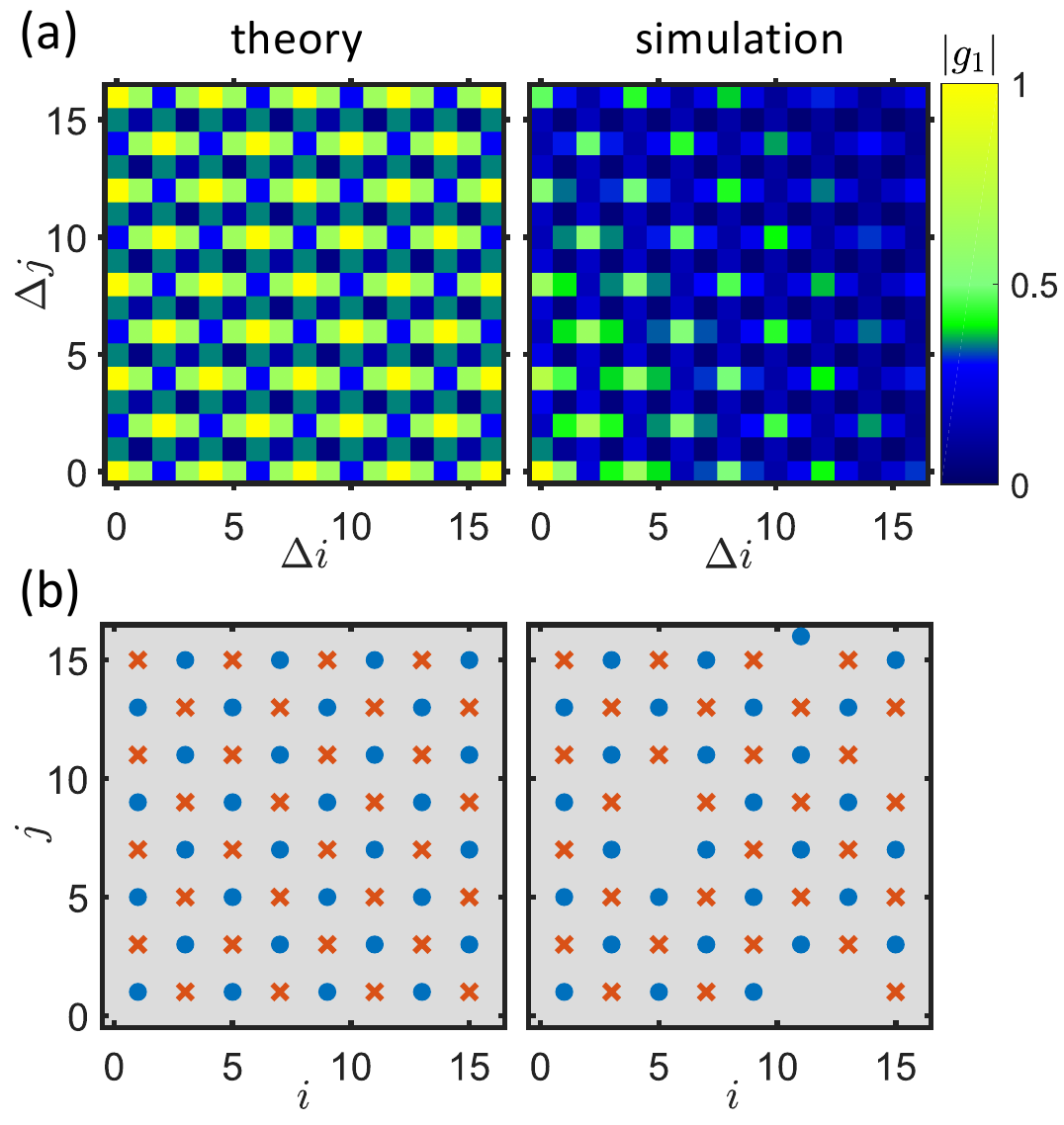}
\caption{Comparison of real-space phase correlation function
and vortex structure between predicted Floquet-Hofstadter ground state (left) and
final dynamically evolved state from simulations (right) at $t=6000\,\mathcal{T}$.
(a)~Absolute value of phase correlation functions
given by
$g_1(\Delta{\bf{r}})=\langle e^{\imath \phi(i,j)}e^{-\imath\phi(i+\Delta i,j+\Delta j)}\rangle$, where $\langle\dots\rangle$ denote averaging over different ensembles and different $(i,j)$ positions with fixed relative displacement $\Delta{\bf{r}}=(\Delta i,\Delta j)a$. Here
$\phi(i,j)$ is the local phase of the wavefunction at ${\bf{r}}=(x,y)=(i,j)a$.
This shows a finite spatial correlation length.
(b)~Distribution of vortices (blue dots) and antivortices (red crosses).
In the simulations, the checkerboard arrangements are present in both the distribution of the phase correlations and that of vortices. These are the predicted signatures of the Hofstadter BEC. }
\label{fig:vornew}
\end{figure}

\begin{figure}[h]
\includegraphics[width=.45\textwidth]
{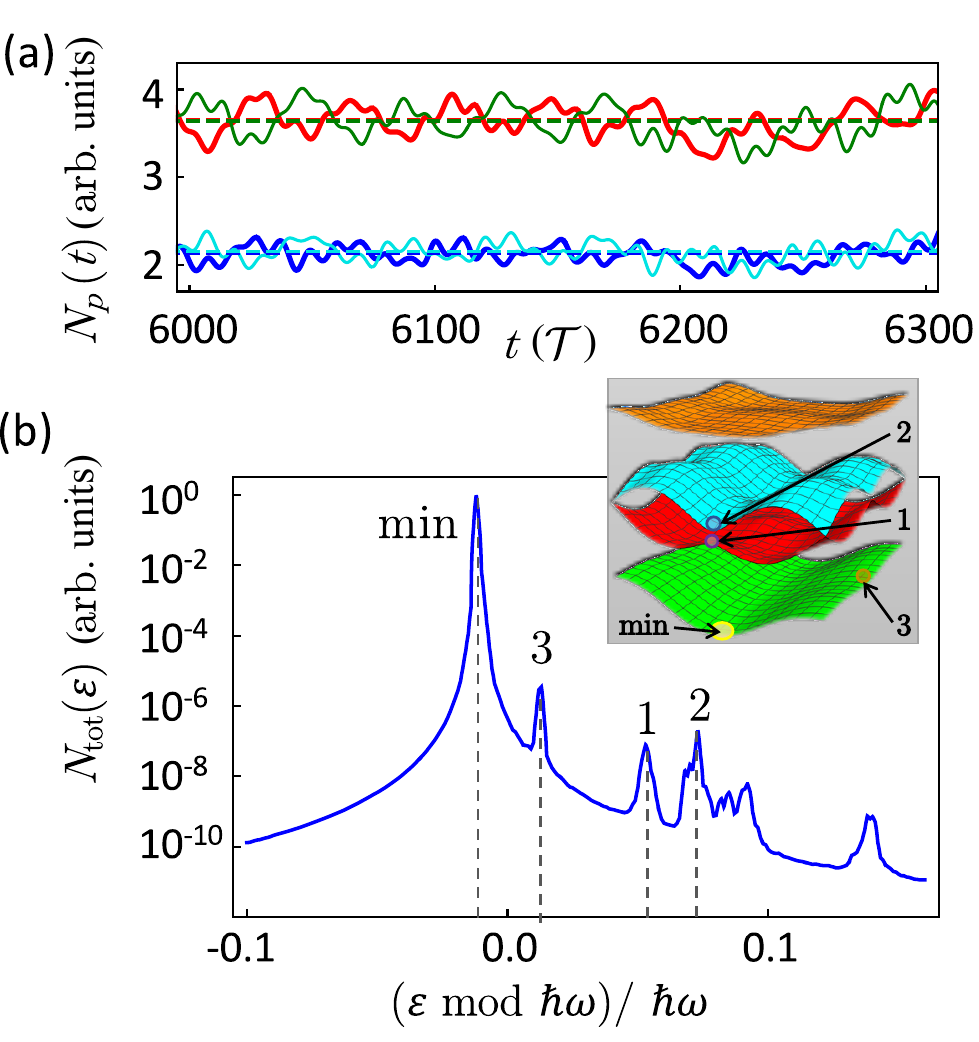}
\caption{(a)~Stroboscopic plot of particle population $N_{\bf{p}}(t)$ at
four characteristic momentum spots of the Floquet-Hofstadter ground state: ${\bf{p}}=(-\pi/2a,0)$ (dark blue: thick lower solid line), $(0,\pi/2a)$ (red: thick upper solid line), $(\pi/2a,0)$ (light blue: thin lower solid line), and $(0,-\pi/2a)$ (green: thin upper solid line). These spots correspond to the red dots in the inset of Fig. \ref{fig:ktime}(a).
We define $N_{\bf{p}}(t)=|\psi_{\bf{p}}(t)|^2$ where $\psi_{\bf{p}}$ is the wavefunction expansion at momentum $\bf{p}$. 
Dashed lines show the average values of the corresponding population curve with the same color. On average, the two spots along the $y$ (or $x$) axis are approximately equally occupied. The population of the two vertical spots
is $1.70$ times that of the two horizontal spots, close to the predicted value 1.67.
(b)~Power spectrum of the wavefunction summed over the four characteristic spots  $N_{\textrm{tot}}(\varepsilon)=\sum_{\bf{p}}|\psi_{\bf{p}}(\varepsilon)|^2$, where $\psi_{\bf{p}}(\varepsilon)$ is the Fourier transform of $\psi_{\bf{p}}(t)$ from $t=5900$ to
$6900\,\mathcal{T}$. At this time the system appears to reach a dynamical equilibrium. The highest peak is normalized to unity.
The lower peaks (labeled 1, 2, 3) show weak occupation of excited states whose quasimomenta are indicated in the inset.
}
\label{fig:Np}
\end{figure}

\section{Evidence for Condensation}

Motivated by these suggestions of condensation in momentum space, and recognizing that these
sorts of analyses are complicated,
we turn to more direct evidence
for a condensate through studies of spatial phase coherence. We wait for time $6000\,\mathcal{T}$ when the atomic population is fully settled into a steady state
in  momentum space.
We then evaluate the phase correlation in real space,
and observe
a long-range phase coherence which extends over 10 sites, see Fig.~\ref{fig:vornew}(a).
This
provides additional and more direct evidence for a BEC
in the Floquet-Hofstadter ground state (referred to as a ``Hofstadter BEC").
At the same time we observe a well organized distribution of vortices
and antivortices. This distribution displays a checkerboard pattern which matches theoretical
predictions determined from the ground state. Moreover, this
vortex checkerboard pattern
\footnote{These vortices are intrinsic and should be contrasted with
vortex patterns
associated with a rotating trap as in Phys. Rev. Lett. 92, 020403.}
is associated with a similar correlation length, see Fig.~\ref{fig:vornew}(b).

Following this strong evidence for a Hofstadter BEC, we next investigate in which Floquet bands the condensate resides. To this end, we study
the time-dependent wavefunctions at the four characteristic momenta  which appear as the red spots in the inset of
Fig.~\ref{fig:ktime}(a).
The populations associated with these four spots exhibit oscillatory behavior when
viewed in a stroboscopic fashion \footnote{The time slice within each cycle is chosen to be at zero phase such that $t$ in Eq.~\eqref{eq:V} is $0+n\mathcal{T}$ where $n$ are integer multiples.}, see Fig.~\ref{fig:Np}(a).
Moreover, the
time-averaged populations
agree well with predictions \cite{Appendix} derived from the ground state in the lowest Floquet-Hofstadter band.

These time-dependent oscillations reveal coherent superpositions involving occupations of excited states.
We are able to extract the energy spectrum of these excitations from the Fourier transform of the wavefunctions in the time domain,
see Fig.~\ref{fig:Np}(b). All Fourier spectra show the same set of peaks consisting dominantly of the ground state with a few excited states. Comparing with the Floquet-Hofstadter band structure, we can identify three excited states from the energies of the weaker peaks \cite{Appendix}, see the inset of Fig.~\ref{fig:Np}(b). We speculate that, these
excitations likely relate to the finite coherence length and defects seen in
Figs.~\ref{fig:vornew}(a)
and
\ref{fig:vornew}(b); presumably they arise
from
the nonadiabatic dynamics in the evolution to the Floquet-Hofstadter ground state.

\section{Dynamics of Hofstadter BEC formation}

While the preceding sections have summarized our central results,
it is useful to understand in more detail the dynamics we observe in our GP simulations.
Particularly noteworthy in this regard is the very novel intermediate heating stage.

\subsection{Numerical details of the Gross Pitaevskii simulations}


The numerical code we use employs~\cite{Andreas}  GPU-based parallel computing and is designed such that it conserves the particle number. The general GP equation is:
\begin{equation}\label{eq:GP}
\begin{aligned}
\imath\hbar\partial_t\psi({\bf{r}},t)=&e^{\imath\gamma}\left[-\frac{\hbar^2\nabla^2}{2m}-\mu+V+g_{int}|\psi({\bf{r}},t)|^2\right]\psi({\bf{r}},t),
\end{aligned}
\end{equation}
where the damping constant
$\gamma$ is set to zero so that our simulations are dissipationless.  Here $\mu$ (set to unity) is the chemical potential, $V=V_{st}+V_{os}$ is the total potential term (see Eq.~\eqref{eq:V} for $V_{st}$ and $V_{os}$),
and $g_{int}$ is the interaction strength which determines the interaction energy $U_0=g_{int}n_0$.
Here $n_0$ is the mean particle density.
Since $U_0$ is directly tuned in our simulations, we focus on this parameter instead of $g_{int}$.

\subsection{ Magnetic-Brillouin-zone entropy}

\begin{figure*}
\includegraphics[width=.95\textwidth]
{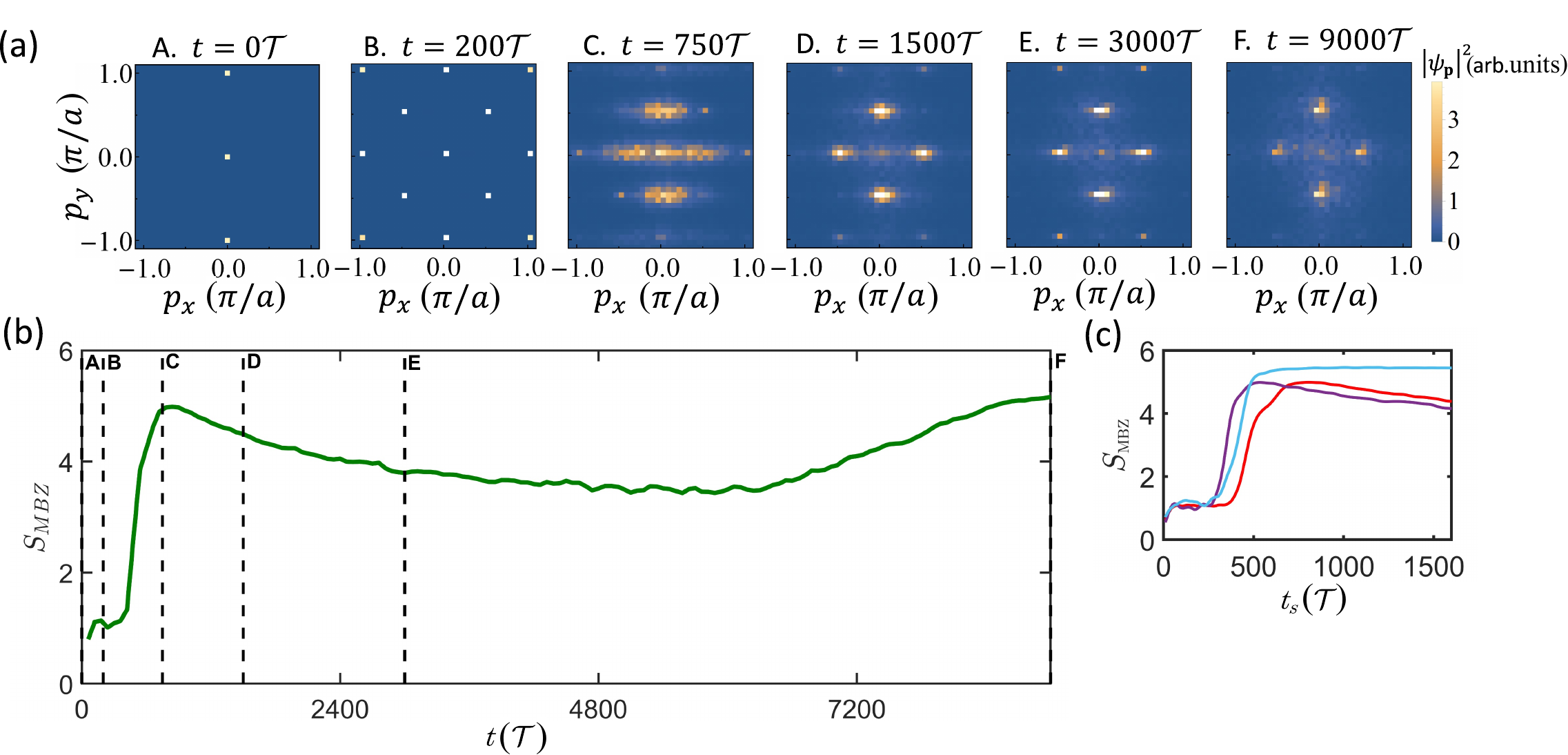}
\caption{\textbf{Evolution of the Floquet system. } (a)~Time-dependent momentum-space distribution of the driven system obtained from simulations. (b)~Entropy as a function of time. Several different time frames are labeled by using the same letters as in panel (a). Note that the longer-time heating, which is associated with an upturn in the entropy,
is evident from panel (b). (c)~The scaling behavior of the entropy with different interaction strengths: $U_0 = 4.5\times 10^{-3} E_R$ (blue), $2.08\times 10^{-3}E_R$ (purple), and $7.5\times 10^{-4}E_R$
(red). At $t=1500\,\mathcal{T}$, the sequence of lines from top to bottom is blue, red, and purple. The scaled time $t_s$ is calculated in such a way that $t_s\propto 1/\sqrt{U_0}$ and $t_s=t$
for $U_0 = 7.5\times 10^{-4}E_R $.}
\label{fig:stime}
\end{figure*}

To more quantitatively characterize these dynamics, in addition to the population
curves shown in Fig.~\ref{fig:ktime},
we introduce an effective time-dependent ``entropy" $S_{\scriptscriptstyle{\mathrm{MBZ}}}$ calculated using states in the first magnetic Brillouin zone. This
serves to quantify the disorder in the momentum distribution, and is defined by
the occupation probability associated with different ${\bf{p}}$ (momentum) spots.
We caution that this ``entropy"
relates to how widely the particle
distribution spreads in momentum space. This does not represent
a thermodynamical definition of entropy.
It serves to describe the sharpness of the momentum peaks over time.
As such this ``entropy"
can, at intermediate times, decrease with time. Eventually, however,
as in Fig.~\ref{fig:stime} the system enters the long time
heating period where the ``entropy" monotonically increases.

\begin{figure*}
\includegraphics[width=.75\textwidth]
{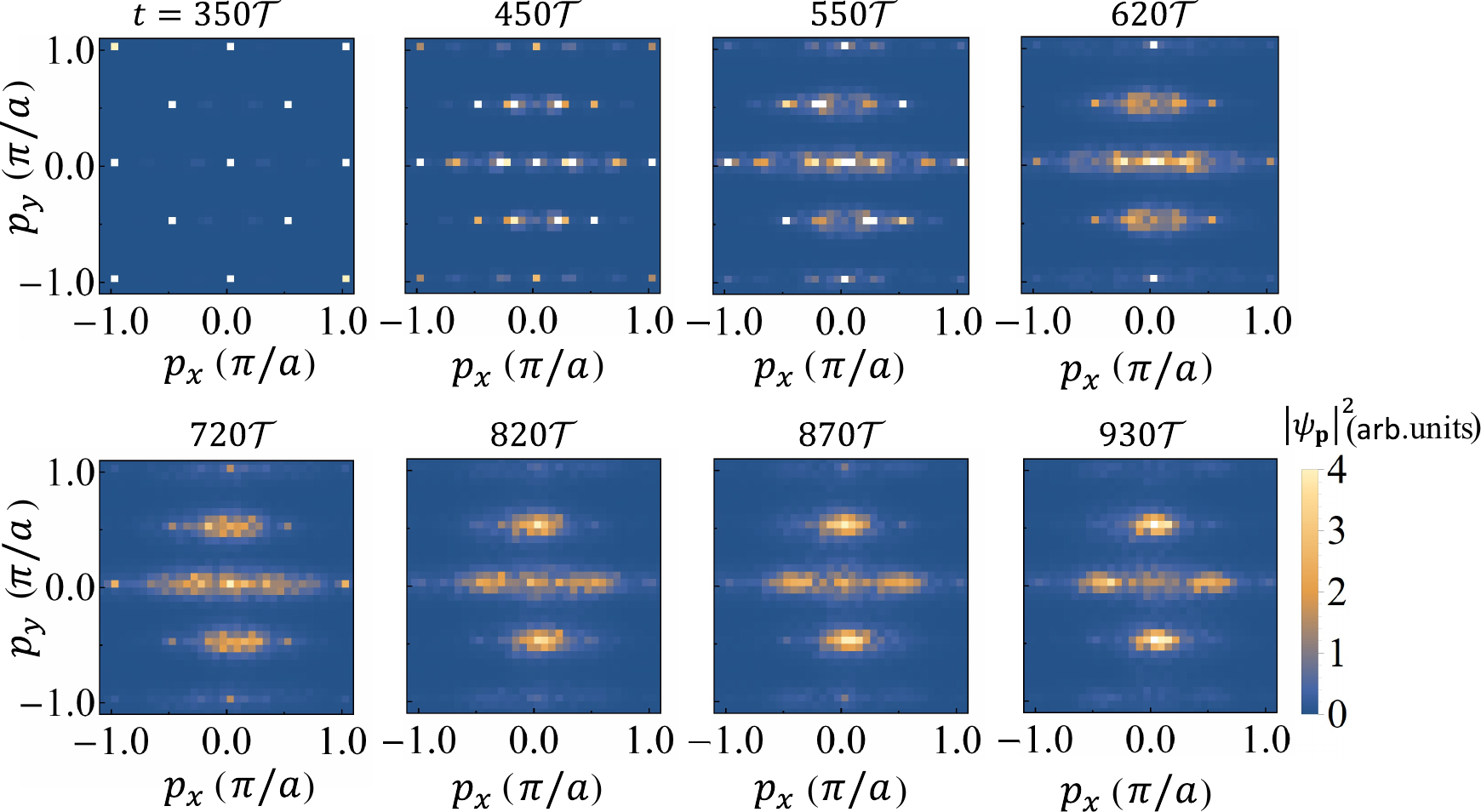}
\caption{\textbf{Evolution of the $\bf{p}$-space distribution within and near the intermediate heating stage.} At the beginning of the intermediate ``heating" stage we
see that the momentum- ($\bf{p}$-) space distribution forms streaks in the horizontal direction; these clear up to form sharp spots at later times when
condensation into the Floquet engineered Hofstadter BEC begins.
}
\label{fig:turbulent}
\end{figure*}


\begin{figure*}
\includegraphics[width=.85\textwidth]
{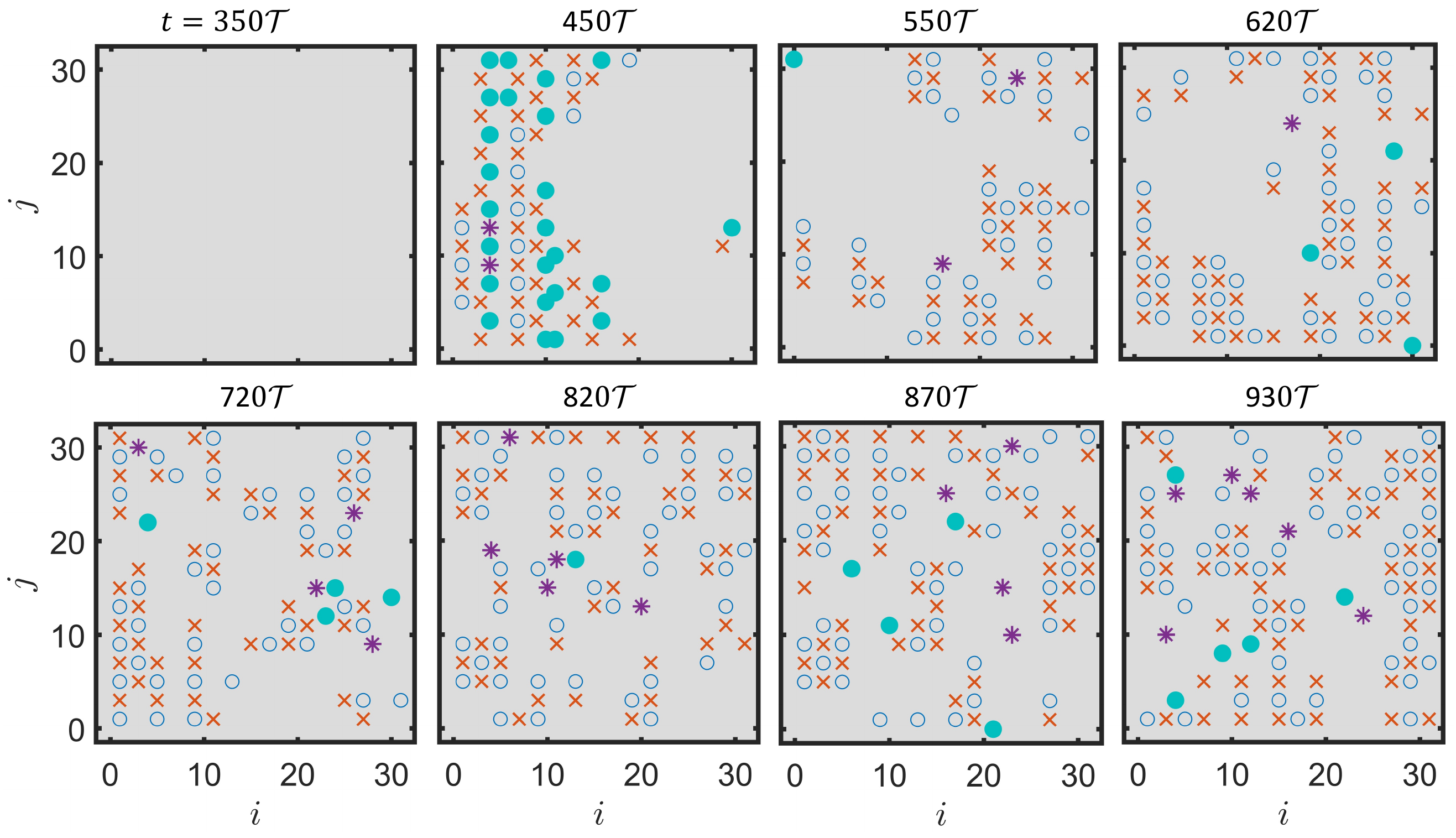}
\caption{\textbf{Time evolution of the vortex distribution within and
near the intermediate heating stage}. The time periods here are the same
as in Fig.~\ref{fig:turbulent}. This figure illustrates the evolution of the wavefunction
phase. The blue circles (red crosses) represent the vortices (antivortices) located at positions consistent with predictions based on the Floquet-Hofstadter ground state, while the filled cyan circles (purple stars) indicate the vortex (antivortex) dislocations.}
\label{fig:vordis}
\end{figure*}


We define
\begin{equation}
S_{\scriptscriptstyle{\mathrm{M B Z}}}=-\sum_{\bf{p}}\rho({\bf{p}}) \ln \rho({\bf{p}})
\end{equation}
where $\rho({\bf{p}})=N_{\bf{p}}/N_t$ is the ratio of the particle number $N_{\bf{p}}$ at momentum $\bf{p}$ to the total number $N_t$. Consistent with the $\bf{p}$-space evolution in Fig.~\ref{fig:stime}(a), we see that the entropy change is also clearly divided into three stages, see Fig.~\ref{fig:stime}(b).
Soon after the initial stage where the entropy is relatively stable, the entropy enters a `heating' phase
where it  exhibits a rapid,
exponential-like growth. This
is associated with a clear maximum in the entropy $S_{\scriptscriptstyle{\mathrm{MBZ}}}$.
We believe this rapid growth
is rather generic, as we have seen it in
simulating other simpler Floquet systems, where it has
been associated with an ``inflaton" picture \cite{inflaton}.
This picture inverts the usual Bogoliubov description
of the excitation spectrum, to describe a collection
of selectively amplified momentum modes which are at lower energy than the
$p=0$ initial (unstable) state.

Interaction effects drive this behavior.
The analysis of growth exponents in simpler systems
\cite{inflaton} suggests a universality
where
the characteristic time scales in the second stage vary as $\propto 1/\sqrt{U_0}$.
With this in mind, Fig.~\ref{fig:stime}(c) shows the
Floquet entropy
presented in terms of rescaled time variables for three different values of
the interaction energy $U_0$.
This scaling with $\sqrt{U_0}$ provides an adequate but imperfect
fit to the power-law dependence in the scaling.
Notably this is consistent with an ``inflaton" model described elsewhere \cite{inflaton}.

We find the system appears to reach dynamical equilibrium in the third
stage near $t=6000\,\mathcal{T}$. The entropy value in this time domain is
rather stable; nevertheless, after $t=6000\,\mathcal{T}$,
$S_{\scriptscriptstyle{\mathrm{MBZ}}}$
begins to slowly increase. This can be interpreted as heating in the long-time limit, which is also expected
to occur experimentally.

\subsection{Analysis of the intermediate heating stage }

We would like to draw particular attention to the
intermediate ``heating" stage we observe. This
represents a crucial (albeit, transient) step in the evolutionary dynamics in which there
appears to be chaotic behavior, as seen from Fig.~\ref{fig:ktime}.
In this section we focus on this behavior by tracking specific features in
the evolution of the system through a sequence of figures.
Whether this chaotic state represents true ``turbulence" or not, it should be noted that the GP
dynamics is associated with weak quantum turbulent behavior \cite{turbulencereview} in non-equilibrated
systems when a persistent source of energy is applied, along with some degree of intrinsic or inevitable dissipation and many-body interactions.

Fig.~\ref{fig:turbulent}
illustrates how the momentum space
distribution of the condensate wavefunction evolves within and near the intermediate heating stage.
The dispersing or spreading out of the momentum peaks suggests highly chaotic behavior
which is observed over an extended time period.
Here the characteristic momentum peaks exhibit streaks along the horizontal direction,
beginning around
$450\,\mathcal{T}$
and persisting for approximately another $300\,\mathcal{T}$.
After this, new momentum peaks
associated with the new Hofstadter condensate appear.
It should be noted that
the asymmetry between the $x$ and $y$ directions which leads to the streaks, reflects the gauge used to implement the artificial vector potential.

Even in our more detailed numerical studies \cite{Appendix} in which high momentum states are filtered out of the GP numerics,
where we see a very ``purified" dynamical evolution, we always find an intermediate chaotic
heating stage.
Strikingly here the numerical filtering
(associated with higher energy band occupation)
is able to remove most of the disorder from our momentum and real space plots,
except during this chaotic evolutionary stage.

Our numerical simulations enable us to more systematically investigate the dynamics to determine how the system effects the transition
from a conventional condensate to one with the highly complex phase pattern of the Hofstadter
BEC. We saw in Fig.~\ref{fig:vornew} that this introduction of phase is reflected in a checkerboard vortex-antivortex pattern.
We now exploit this pattern to probe
how phase coherence is dynamically established. This is illustrated
in Fig.~\ref{fig:vordis}
which indicates in cyan and purple
how the vortex and anti-vortex dislocations evolve against the background checkerboard pattern.
The time sequence is the same as for Fig.~\ref{fig:turbulent}.
What is striking is the suddeness of flux penetration.
These vortex-antivortex pairs with a large fraction of dislocations
initially appear
precipitously at
$450\,\mathcal{T}$. They then rapidly reorganize as the dislocations
are removed and the extended checkerboard pattern is systematically developed.
Interestingly, this time frame where flux abruptly penetrates is roughly
the same as the onset of spreading out of
sharp momentum peaks found in the initial condensate.

\section{Conclusions}


In conclusion, the work in this paper addresses the important and complicated question
of how one can successfully guide a wide class of Floquet engineered systems
\cite{Dalibard_2014,Monika_2018,Shuai} into the quantum regime. From our simulations, we show that a dynamical
conversion of a regular BEC into the Hofstadter ground state can be realized with high efficiency. The dynamics
involves an intriguing chaotic ``heating" stage during which ``magnetic" flux rapidly penetrates.

More generally, this paper addresses a need in the quantum gas community to prepare novel quantum matter with
Floquet engineering. Concerns about heating are widespread not only for dynamical preparation of topological
matter \footnote{Alternative schemes, besides Floquet engineering, have
recently been discussed in the literature \cite{Erich,Liu}, which may introduce some form of heating as well.} but more generally to surmount barriers \cite{DeMarco} for reaching the quantum regime.
Quite intriguingly,
the system is seen to overcome these
challenges and the way it does so is in many ways reminiscent of evolution in cosmological models;
this involves a
similar time progression including an intermediate turbulence \cite{reheating} en route to equilibration.


\vskip1mm
\section*{Acknowledgements}
We acknowledge L. Feng, L. W. Clark, Weihan
Hsiao, B. M. Anderson, Nigel Cooper, T. Bilitewski, I. S. Aronson, and A.
Polkovnikov for helpful discussions. We acknowledge support by the U. S. Department of Energy, Office of Basic Energy Sciences, under contract number DE-SC0019216, the Army Research Office under Grant No. W911NF-15-1-0113, and the University of Chicago Materials Research Science and Engineering Center, funded by the National Science Foundation under Grant No. DMR-1420709. A. Glatz acknowledges support from the U. S. Department of Energy, Office of Science, Basic Energy Sciences, Materials Sciences and Engineering Division. F. Setiawan acknowledges support from the Army Research Office under Grant No. W911NF-19-1-0328.

\appendix
\numberwithin{equation}{section}



\section{Simulating the ideal Hofstadter model by Floquet-engineered Hamiltonian}

\begin{figure*}[h]
\includegraphics[width=.8\textwidth]
{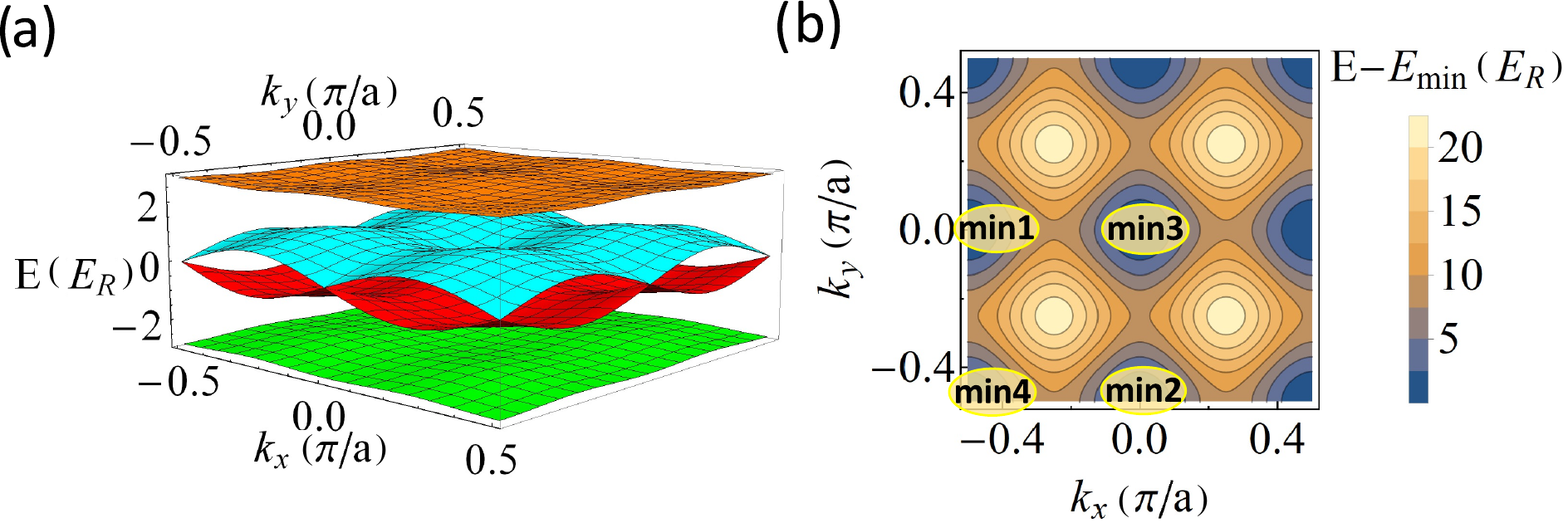}
\caption{\textbf{Band structure and wavefunction for the ideal Hofstadter Hamiltonian.} (a)~3D plot of the band structure for the tunneling parameters $J_x^\prime=J_y^\prime=1 E_R$. (b)~Color contour plot of the lowest band. There are four different degenerate ground states labeled as min1 at ${\bf{k}}=(-\pi/2a,0)$, min2 at ${\bf{k}}=(0,-\pi/2a)$, min3 at ${\bf{k}}=(0,0)$, and min4 at ${\bf{k}}=(-\pi/2a,-\pi/2a)$. 
}
\label{fig:iHH}
\end{figure*}

The general Harper-Hofstadter model we simulate with Floquet engineering is
\begin{align}
H_{iHH}=&-\sum_{i,j}\left(J_x^\prime e^{\imath\phi_{i,j}^x}\hat{a}^\dagger_{i+1,j}
\hat{a}_{i,j}+J_y^\prime e^{\imath\phi^y_{i,j}}\hat{a}^\dagger_{i,j+1}\hat{a}_{i,j}+\mathrm{h.c.}\right),
\end{align}\label{eq:tbind}
which can be approached using the Floquet Hamiltonian [with the lattice potential given by Eq.~\eqref{eq:V} of the main text] in the ideal limit of $J^\prime_x/\hbar\omega\rightarrow 0$ and $J^\prime_y/\hbar\omega\rightarrow 0$. Here, the tunneling phases in $x$ and $y$ directions $\phi_{i,j}^{x}$ and $\phi_{i,j}^y$ are associated with the applied vector field in each direction. The same filling factor can be associated with different gauges \cite{Ketterle2}. To calculate the corresponding band structure in a chosen gauge (used in Ref. \onlinecite{Cooper}), we use the magnetic translation operator to identify the eigenstates \cite{Monika_thesis}. The results are shown in Fig.~\ref{fig:iHH}, where panel (a) shows the 3D band structure, and panel (b) shows the 2D color contour plot of the lowest band.
There are four different ground states in the lowest band. This is to be contrasted with the unique ground state of the Floquet-engineered Hamiltonian realized by implementing Eq.~\eqref{eq:V} with a moderately large modulation frequency. The four-fold degeneracy in the ground states of the ideal Hofstadter model is lifted in the Floquet-engineered case by higher-order terms in $J^\prime_{x}/\hbar\omega$ and $J^\prime_y/\hbar\omega$ which hybridize states connected by wavevectors introduced by the oscillating lattice.

\begin{figure*}[h]
\includegraphics[width=.72\textwidth]
{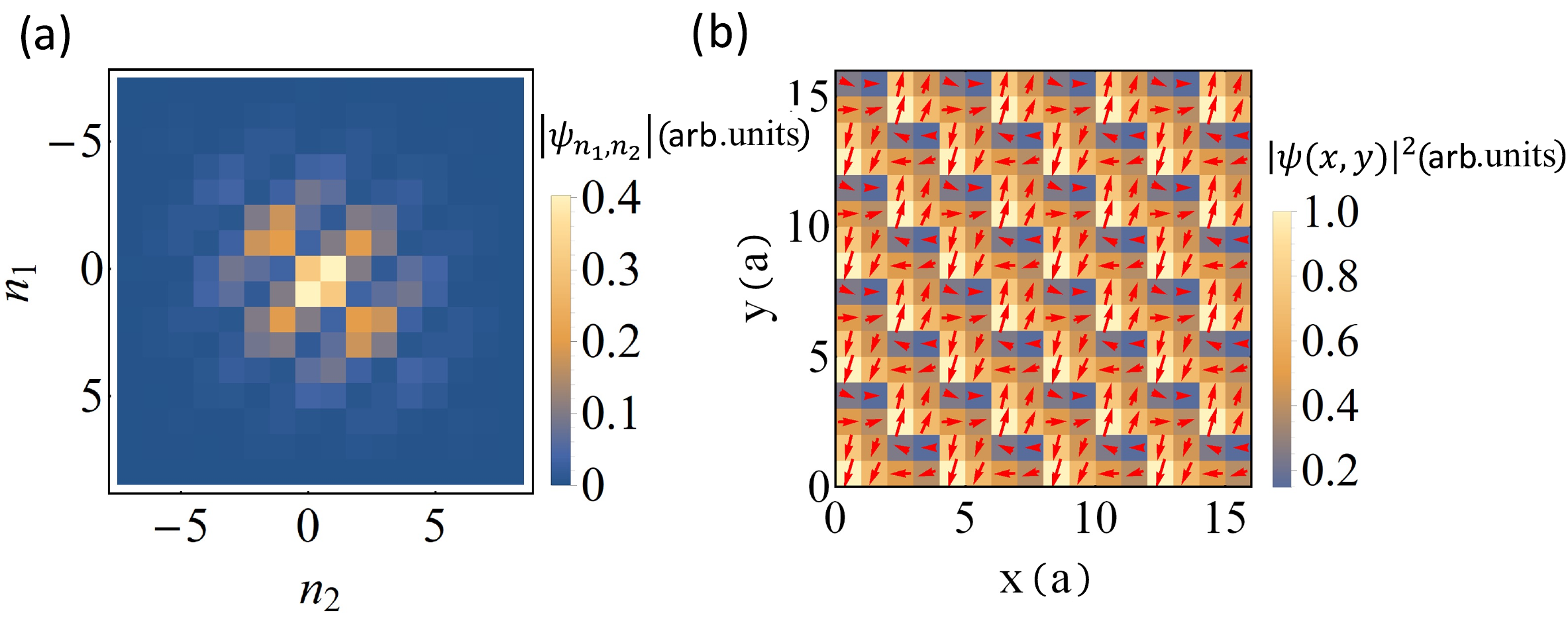}
\caption{\textbf{Ground-state wavefunction of the Floquet band.} (a)~Momentum-space expansion of the ground state in terms of the reciprocal vectors ${\bf{G}}=n_1{\bf{G_1}}+n_2{\bf{G_2}}$. (b)~Real-space representation of the wavefunction showing both phase and
amplitude. Note that both the phase and amplitude distributions display a checkerboard pattern. }
\label{fig:fground}
\end{figure*}
We now present more details on the band structure predicted by Floquet theory. As shown by Fig.~\ref{fig:bandcross} in the main text, there is only one unique ground state corresponding to two equivalent degenerate states in the lowest band. The ground state can be expanded in terms of the total momentum ${\bf{p}}={\bf{k}}+{\bf{G}}$ at ${\bf{k}}=(-\frac{\pi}{2a},0)$. The corresponding amplitude of the wavefunction at different $\bf{G}$ is presented in Fig.~\ref{fig:fground}(a) where ${\bf{G}}=n_1 {\bf{G_1}}+n_2{\bf{G_2}}$ with ${\bf{G_{1,2}}}=(\frac{\pi}{2a},\pm\frac{\pi}{2a})$ being wavevectors of the oscillating potential $V_{os}$ and $n_{1,2}$ being integers. The distribution of the ground-state wavefunction has four dominant peaks in $\bf{G}$, which correspond to the four characteristic $\bf{p}$ spots shown by Fig.~\ref{fig:wave} in the main text. The relative phases of the ground-state wavefunction at the four spots are
$\theta_l-\theta_r\approx\pi,\theta_t-\theta_b\approx\pi,\theta_b-\theta_r\approx2.04$ radians, where the subscripts correspond to the left ($l$), right ($r$), top ($t$), and bottom ($b$) momentum spots, respectively.
We present an overlay picture of the density and phase distributions in Fig.~\ref{fig:fground}(b) for the ground-state wavefunction in the Floquet-engineered case. 

\begin{figure*}[h]
\includegraphics[width=.75\textwidth]
{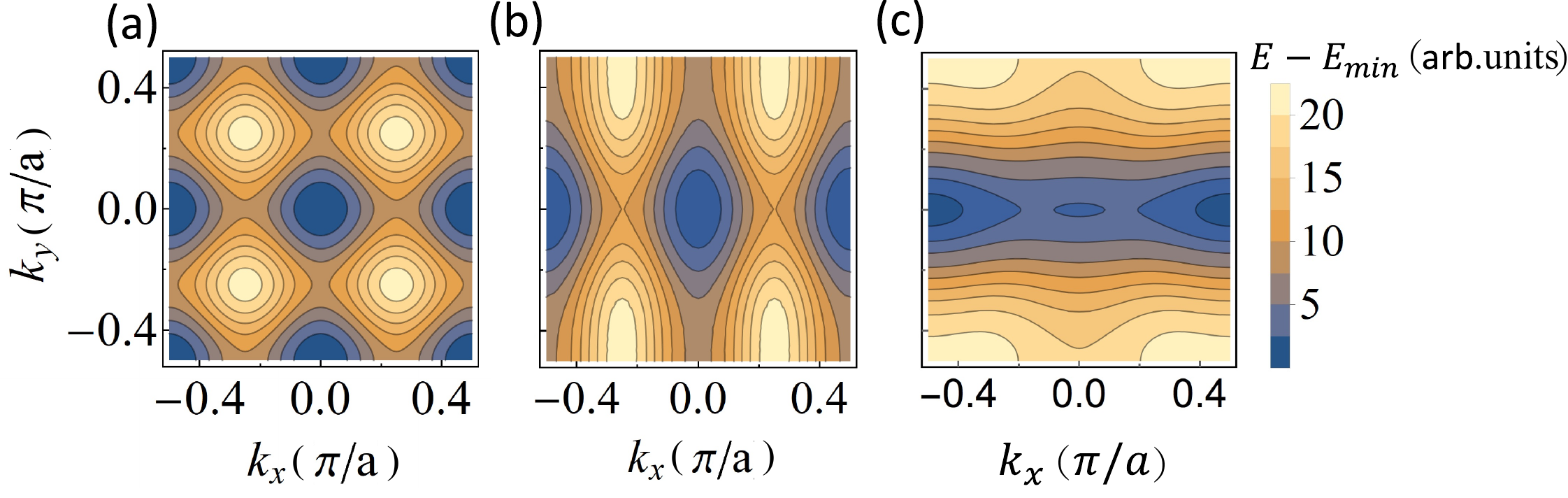}
\caption{\textbf{Effects of increasing the lattice depths} (a)~Lowest-energy band for the ideal Hofstadter Hamiltonian calculated using the parameters in Fig.~\ref{fig:iHH}. (b)~Lowest-energy band for the Floquet Hamiltonian with very deep lattices. Here $V_y=20 E_R,\,V_x=22E_R,\,V_{yl}=1.0625E_R$, $\hbar\omega=0.99664 E_R$, and $\kappa=0.58\hbar\omega$.  (c)~Lowest Floquet band with lattice
depths used in the main text. The energy dispersion is shown for the first magnetic Brillouin zone. And the energy is measured with respect to the lowest energy $E_{min}$. The energy units (denoted by the arbitrary unit [arb]) and $E_{min}$ are different for different cases.   }
\label{fig:iHHcross}
\end{figure*}

\begin{figure*}[h]
\includegraphics[width=.85\textwidth]
{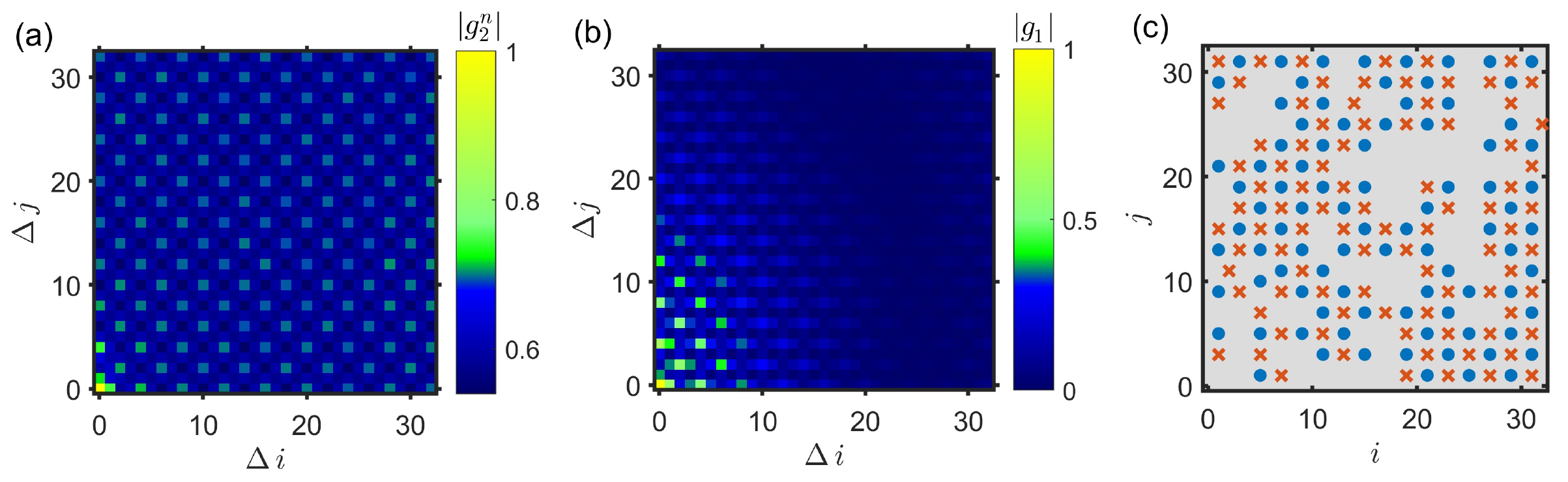}
\caption{\textbf{Real-space wavefunction in larger systems. } (a)Absolute value of density correlation function $g_2^n(\Delta i, \Delta j)=\langle n(i,j)n(i+\Delta i,j+\Delta j)\rangle/\langle n\rangle ^2$ where $n(i,j)=|\psi(i,j)|^2$ is the local density. Here, $\langle\cdots\rangle$ denotes averaging over different ensembles and different $(i,j)$ positions with fixed relative displacement $(\Delta i,\Delta j)$. (b)Absolute value of phase correlation function $g_1$. The definition is given in the main text. (c)~Distribution of vortices (blue dots) and antivortices (red crosses). While the unique ground state yields constant density correlation even at large distance, the phase correlation reveals a finite size for
the physically coherent regions. The vortex structure is associated with vortex defects
rather than dislocations; this reflects the absence of vortices and antivortices.} \label{fig:Scorr}
\end{figure*}
\begin{figure*}[ht]
\includegraphics[width=.85\textwidth]
{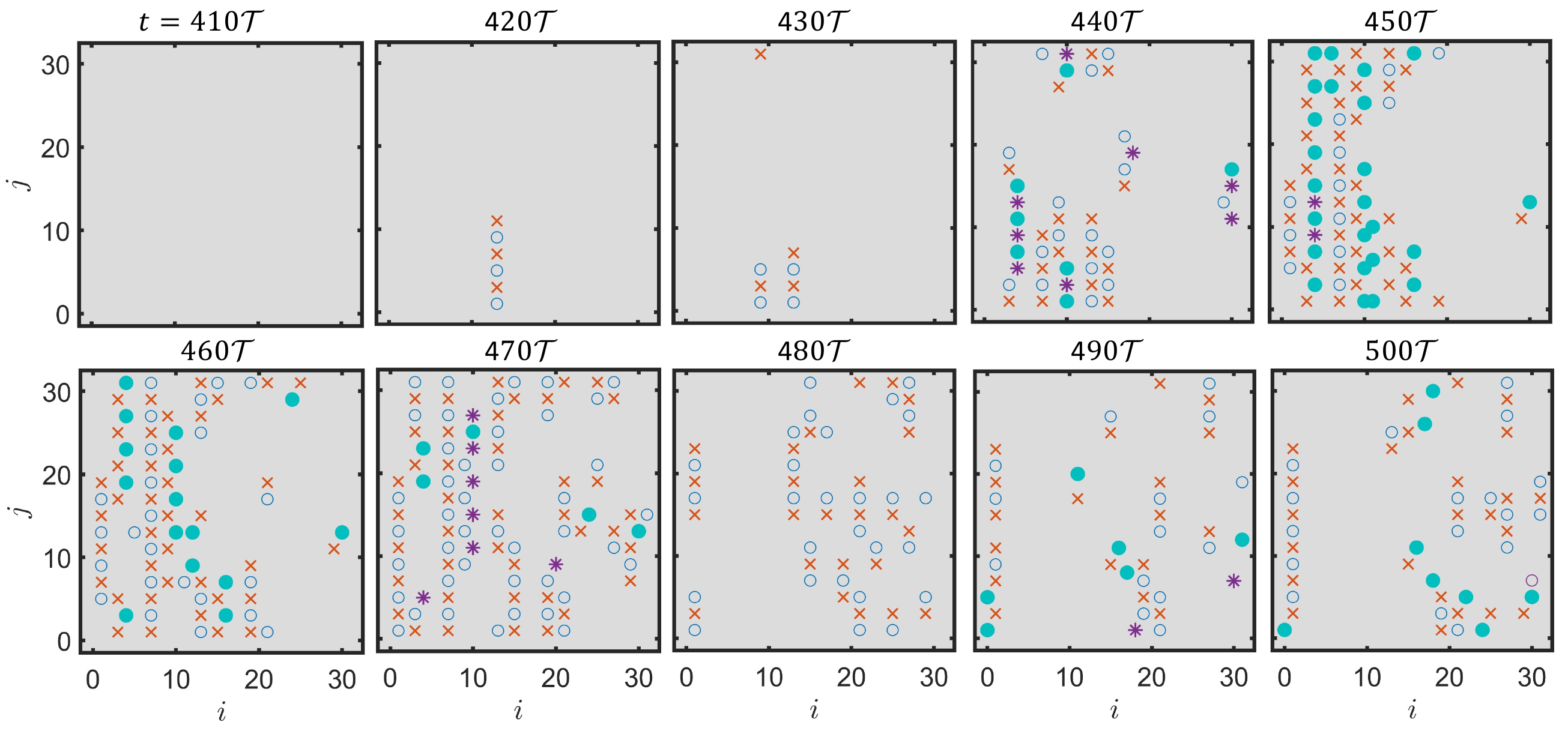}
\caption{Illustration of flux penetration seen at early times
within intermediate heating stage.}
\label{fig:transientturbulent}
\end{figure*}

By tuning the shaking frequency and the lattice depth in the Floquet Hamiltonian, we can either go to the limit of an ideal Hofstadter band or stay with the current Floquet band. In the following, we consider an intermediate case between the two scenarios by proper choice of the parameters. We find
an intermediate state that exhibits two minima in the lowest band, see Fig.~\ref{fig:iHHcross}. We can clearly see a trend of fewer minima in the lowest band with shallower lattices and smaller shaking frequencies. One should note, however, though the minima in Fig.~\ref{fig:iHHcross}(c) seem to be surviving minima in Fig.~\ref{fig:iHHcross}(a) at ${\bf{k}}=(\pm\frac{\pi}{2a},0)$, in fact, the ground states are quite different for these two cases.

\section{Comparison between Floquet prediction and GP simulation results}

\subsection{Comparison of relative phases at the four characteristic $\bf{p}$ spots for the ground state}
In the main text, we have already seen from the simulations that the particle distribution is peaked at four characteristic $\bf{p}$ spots within the magnetic Brillouin zone (see Fig.~\ref{fig:ktime}); this
agrees well with the predictions shown in Fig.~\ref{fig:wave}(d), which are calculated by diagonalizing the Floquet Hamiltonian. To be more quantitative, we look at the ratio between the wavefunction amplitudes at the four spots. The time-averaged populations from the simulations shown by Fig.~\ref{fig:Np}(a) are consistent with our predictions extracted from Fig.~\ref{fig:fground}(a).
The occupation at ${\bf{p}}=(0,\pm\frac{\pi}{2a})$
is $1.70$ times that at ${\bf{p}}=(\pm\frac{\pi}{2a},0)$, close to the predicted value 1.67.

We can also check the relative phases between these characteristic spots. From our simulations, we can directly obtain the wavefunction expansion in $\bf{p}$ space. We find that the relative phases are as follows: $\langle\theta_l-\theta_r\rangle\approx 3.10,\langle\theta_t-\theta_b\rangle\approx  3.15,\langle\theta_b-\theta_r\rangle\approx  1.14$. These are consistent with our prediction presented in Sec. I, except that there is approximately a difference of
order unity in $\langle\theta_b-\theta_r\rangle$. This discrepancy, though quite robust and also present
in our filtered simulations (to be discussed in Appendix C.~2, where the BEC is much cleaner), seems
to be related to interaction effects, as when we decrease the interaction energy to $2.5\times 10^{-4}E_R$, the phase difference increases significantly to 1.44.

\subsection{Comparison of frequency spectrum for both the ground state and excited states}
In Fig.~\ref{fig:Np} of the main text, we have seen that besides the dominant ground state in the observed BEC, there are also excited states.
Table~\ref{tab:states2} presents a summary of the energy comparisons which allow us to identify some of these
excited states.
When we introduce a high-frequency filtration in our GP dynamics (discussed in Appendix C.~2),
we see a negligibly small occupation of these higher bands.

\begin{table}[h]
  \centering
  \caption{\textbf{Identification of the
ground state (min) and 3 excited states.} The table
compares energies from simulations ($\varepsilon_{sim}$) and predictions ($\varepsilon_{pre}$) for the ground state and excited states in the Floquet-Hofstadter band. As in the main text, min denotes the `ground state', while 1, 2, 3 refer to the same states as appear in Fig.~\ref{fig:Np}.}
 $\begin{array}{|c|c|c|c|}
    \hline
      \textrm{Spot index} &\textrm{Band index}& \varepsilon_{sim}\, \textrm{modulo}\, \hbar\omega & \varepsilon_{pre}\,\textrm{modulo}\,\hbar\omega \\
     \hline
     \textrm{min} &1&-0.011&-0.012 \\
      \hline
      1 &2&0.054 & 0.057  \\
      \hline
      2 &3&0.070 &0.068   \\
      \hline
      3 &1&0.014 & 0.013  \\
      \hline
    \end{array}$
  \label{tab:states2}
\end{table}

\subsection{Comparison between the real-space correlation functions and vortex structure}

The BEC we obtain from simulations is a combination of mostly the ground state
and a small number of excited states, which is also reflected in the finite coherence length in real space (see Fig.~\ref{fig:vornew}). To check this more thoroughly, we look at an even bigger system ($4\times4$ times larger). In Fig.~\ref{fig:Scorr}, we show the real-space density and phase correlation functions of the BEC obtained from our GP simulations for such a system.
Indeed, we see defects are present leading to a finite range for the spatial phase coherence. They appear more directly (mostly as vacancies) in the vortex-antivortex distribution.

It is important to note that the positions of the vortices and antivortices in the distribution are not interchangeable as there is a
single, non-degenerate ground state. This can be viewed as a chiral asymmetry associated with an artificial magnetic
field. The defects we observe are manifested as an absence of a vortex or antivortex as distinct from dislocations, and we believe
they derive, at least in part, from the excited states co-existing with the condensate \footnote{We note that, this alternating pattern is also found in the initial magnetic Brillouin zone, but, importantly,
it disappears during a large portion of the evolution time, until well past the end of the inflation stage}.

\section{Dynamics of the Hofstadter BEC formation from GP simulations}

\subsection{Flux penetration dynamics}

In Fig.~\ref{fig:vordis} of the main text we discussed the rather precipitous appearance of magnetic flux. Here in this
appendix we
look at this behavior in a more refined way focusing on the transient
period where the vortices first appear. This is illustrated in
Fig.~\ref{fig:transientturbulent}
which shows how the system rapidly evolves from one state with uniform phase to another with a complex
phase pattern. The color scheme is the same as in Fig.~\ref{fig:vordis} and the cyan (purple)
coloration labels vortex (antivortex) dislocations.
This figure illustrates
the very transient appearance of these dislocations.
Their initial density is very high which presumably represents the onset of
flux penetration; it then rapidly decreases as the checkerboard pattern
of organized vortices associated with the Hofstadter BEC begins to emerge. We speculate that
this transient high density of dislocations might
indicate some degree of turbulent behavior which arises due to the onset of an artificial
vector potential.

\subsection{Removing high-frequency contributions: purifying the BEC}
\begin{figure*}[ht]
\includegraphics[width=.95\textwidth]
{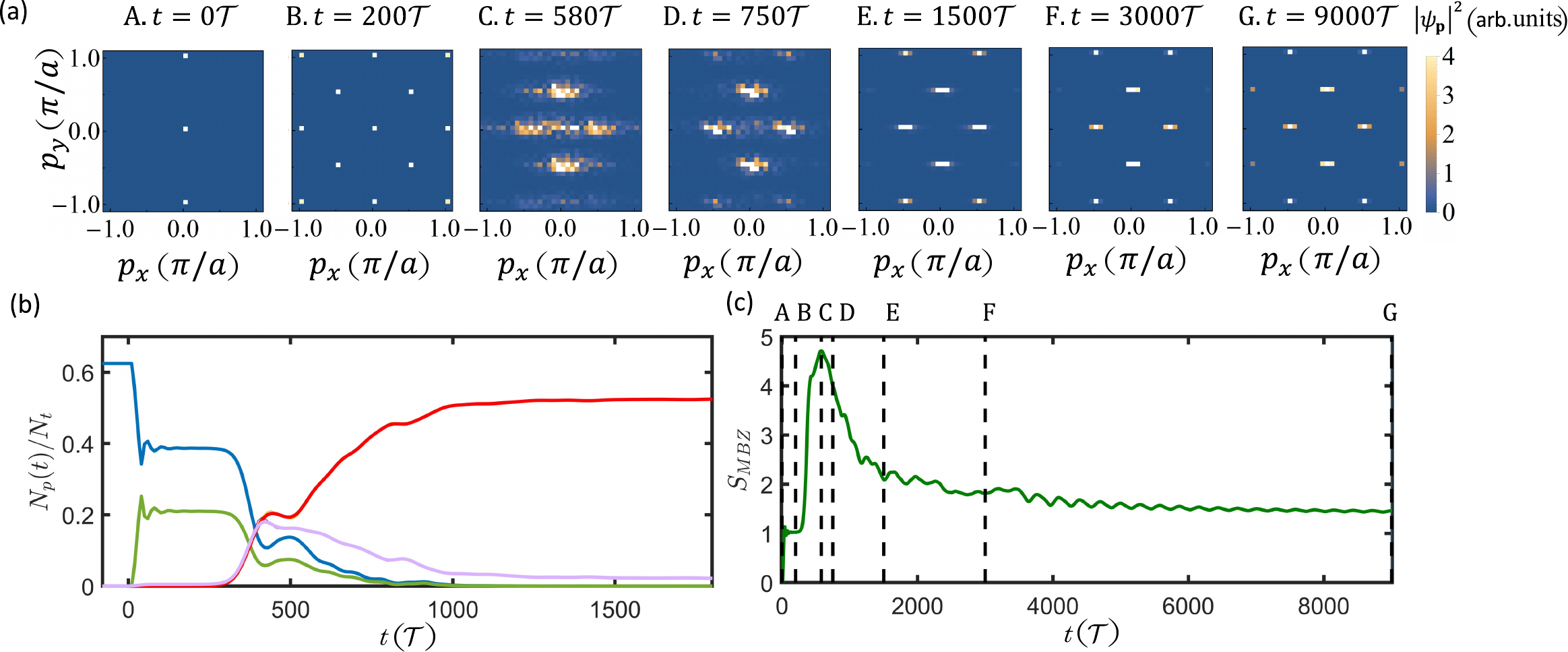}
\caption{\textbf{Effects of high energy filtering in the
simulations}
(a) Density image in $\bf{p}$-space at different times. (b) Population curves,
as defined in Fig.~\ref{fig:ktime}. At $t=500\,\mathcal{T}$, the sequence of lines from top to bottom is red, purple, blue, and green. (c) Entropy as a function of time. We find that with a moderate filtration, the system reaches a cleaner BEC state with the excitations almost
completely gone. Importantly, the intermediate heating stage is still present, and persists for
a shorter period of time. The final condensation fraction is greatly enhanced.
Finally there is no sign of long time heating as seen from
the entropy which does not increase in the long-time limit.}
\label{fig:filter}
\end{figure*}

The dynamics we simulate in our GP equation do not involve energy dissipation, once the initial state is
established \footnote{By starting from a random configuration, the system evolves
to the proper initial equilibrium state for
a negative phase angle $\gamma<0^\circ$ in Eq.~\eqref{eq:GP}, see Ref. \onlinecite{Kathy}.}.
We stress that our simulations without dissipation are consistent with experimental conditions in atomic systems.
We might expect that
in actual experiments, the system picks up some higher momentum ($\bf{p}$) excitations over longer time scales. Indeed, these
are presumably responsible for the final heating stage. High $\bf{p}$
excitations do occur in our simulations. 

It is also informative, then, to compare the behavior of the system when
these high-energy excitations are removed, as is often done when studying the stochastic GP equation \cite{MattDavis}.
For this reason, we apply a high-momentum filter at each time step of the numerical integration, i.e., multiplying the Fourier transform of the order parameter by a Gaussian function. Physically, this process may represent intrinsic losses, such as those due to three-body and other collisions.

When the filtration is weak (where the momentum threshold above which the modes will be removed is high), we find the behavior is generally unaffected except that the higher-band excited states are no longer present,
and the peak structure of the target BEC becomes sharper as shown in
Fig.~\ref{fig:filter}.

This can be seen more clearly through a comparison with the unfiltered case. Panels (a) and (c) in Fig.~\ref{fig:filter} can be contrasted with panels (a) and (b) in Fig.~\ref{fig:stime}, while Fig.~\ref{fig:filter}(b) can be contrasted with Fig.~\ref{fig:ktime}(a). This comparison
reveals those features arising from higher-energy states, presumably deriving from the role
of the higher bands. For the most part the early-time evolution
is similar. The oscillations which are present without filtration are greatly diminished,
thus suggesting that these may come from higher band occupation.
We stress that the chaotic, intermediate heating stage is still
present. We also observe that the longer-time heating (seen in the entropy plot) vanishes with filtration and consequently the BEC
is more stable. Because the presence of this longer-time heating appears more realistic, we conclude that the unfiltered case is the more physical.

\subsection{Role of Kibble-Zurek mechanism}

\begin{figure*}[ht]
\includegraphics[width=.43\textwidth]
{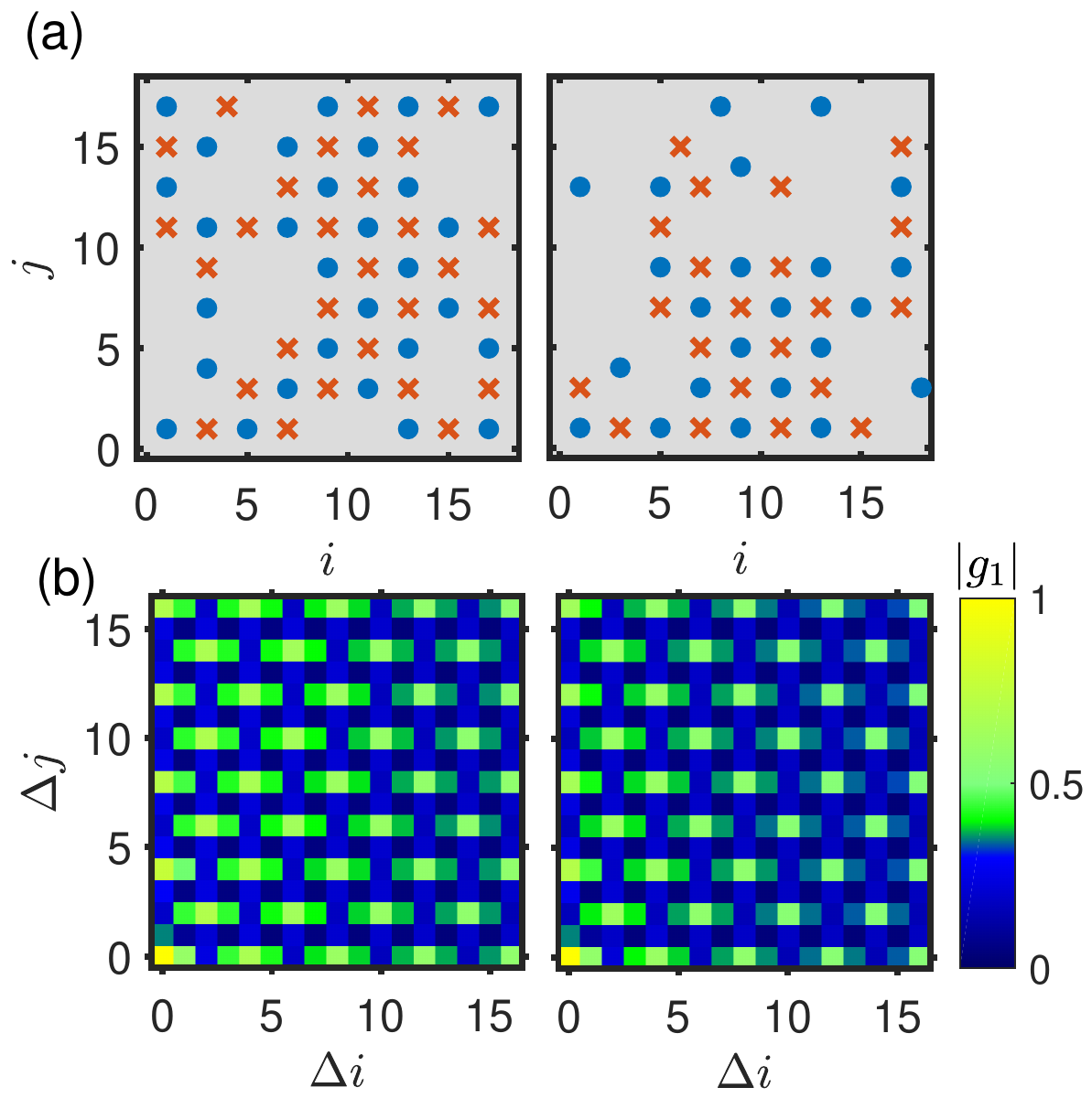}
\caption{\textbf{Comparison of coherence
between systems with different ramping rates at $t=6000\,\mathcal{T}$.} (a)~Vortex (blue dot) and antivortex (red cross) distribution of real space wavefunction. (b)~Absolute value of phase correlation function $g_1$ for real space wavefunction. The ramp time is $300\,\mathcal{T}$ (left panel) and $30\,\mathcal{T}$ (right panel), respectively. } \label{fig:kb}
\end{figure*}

Here, we want to briefly discuss the role of the Kibble-Zurek (KZ) mechanism. When a dynamical system crosses a critical point like a phase transition point by ramping a key parameter such as the shaking amplitude, the correlation length in the ordered phase is determined by the ramping rate. The slower the rate is, the bigger the domain size or the correlation length.  Here, one might wonder if
similar effects are responsible for defects observed in this paper. We have a similar transition point, but the change of the band structure and thus the ground-state wavefunction is abrupt (and more first-order like) at the critical shaking amplitude $\kappa=0$.

Our results suggest an absence of important effects associated with the KZ mechanism. This can be seen by comparing the real-space distribution of the vortices between cases with different ramping rates, see Fig.~\ref{fig:kb}. While the right panel in Fig.~\ref{fig:kb} corresponds to a very fast ramp (essentially a quench), the left panel corresponds to a slow process (with a ramping period 10 times as long). The domain size and phase correlation length are found to be comparable, indicating the lack of
important KZ effects here.
\begin{figure*}[ht]
\includegraphics[width=.86\textwidth]
{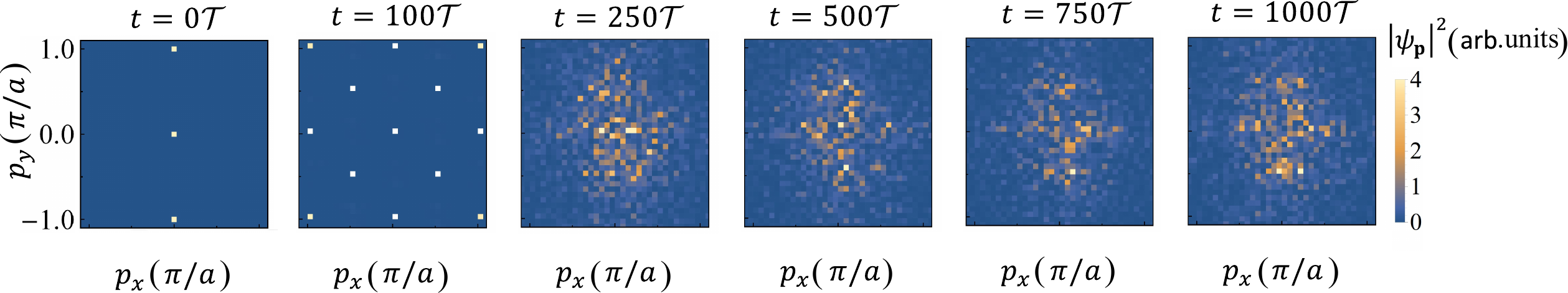}
\caption{\textbf{Evolution of the $p$-space distribution for stronger inter-particle
interactions: $U_0 = 0.009E_R$}. Here we increase the repulsion by a factor of 10 compared to that used in the main text. We see that the system seems to never reach the Hofstadter BEC and remains in a chaotic state throughout the simulation. }
\label{fig:ksmearing}
\end{figure*}

\subsection{Strong interaction effect: absence of condensation}

It is important to investigate the effects associated with the interaction strength $U_0$, since the validity of
Floquet predictions is based on assuming that such interactions are negligible. GP simulations allow the simultaneous
incorporation of Floquet
engineering along with interaction effects. Our results show that
with a moderately large
$U_0$
the evolutionary behavior tends to be very noisy without
clearing up, see Fig.~\ref{fig:ksmearing}. This behavior suggests the failure to form a BEC. Indeed this is
consistent with observations in Ref. \onlinecite{Ketterle}, where, when the collision rate is too high,
this is seen to seriously disturb the single-particle band structure.

\end{document}